\documentclass[12pt,preprint]{aastex}

\newcommand\bb[1] {   \mbox{\boldmath{$#1$}}  }

\newcommand\del{\bb{\nabla}}
\newcommand\bcdot{\bb{\cdot}}
\newcommand\btimes{\bb{\times}}

\newcommand\xb{ {X_b} }
\newcommand\yb{ {Y_b} }
\newcommand\zb{ {Z_b} }

\def\dd{\partial}

\def\beq{ \begin{equation} }
\def\eeq{ \end{equation} }
\def\spose#1{\hbox to 0pt{#1\hss}} 
\def\ltsim{\mathrel{\spose{\lower.5ex\hbox{$\mathchar"218$}}
\raise.4ex\hbox{$\mathchar"13C$}}}

\slugcomment{To appear in Ap.~J.}

\shorttitle{Viscous Shear Instability}
\shortauthors{Balbus}

\begin{document}


\title{Viscous Shear Instability\\
    in Weakly Magnetized, Dilute Plasmas}


\author{Steven A. Balbus\altaffilmark{1}}
\affil{\'Ecole Normale Sup\'erieure, Laboratoire de Radioastronomie,
24, rue Lhomond,\\
75231 Paris, France}

\email{sb@virginia.edu}


\altaffiltext{1}{permanent address: VITA, Astronomy Department, University of Virginia,
Charlottesville, VA 22903}

\begin{abstract} When the ion mean free path much exceeds the Larmor
radius in a plasma, the viscous stress tensor is altered dramatically, and
depends only upon quantities measured along the field lines.  This regime
corresponds to typical interstellar medium conditions in galaxies and
protogalaxies, even if the magnetic field is extremely weak, with a
negligible Lorentz force on all scales of interest.
In this work, the only role of the magnetic field is to channel angular
momentum transport along its lines of force. We show that differential
rotation in such a gas is highly unstable, with a maximum growth rate
exceeding that of the magnetorotational instability.  The regime
of interest has been treated previously by plasma kinetic methods.
Where there is overlap, our work appears to be in agreement with the
kinetic results.  The nonlinear outcome of this instability is likely to
be a turbulent process, significantly augmenting the magnetorotational
instability, and important to the initial phases of the amplification
of small galactic magnetic fields.

\end{abstract}



\section{Introduction}

The magnetorotational instability, or MRI, has become central to our
understanding of turbulent angular momentum transport in accretion disks
(e.g. Balbus 2003).  The instability is important even when (indeed,
especially when) the magnetic energy density is small compared with the
thermal energy density.  It is this feature, the fact that the stability
of the gas is hypersensitive to the presence of subthermal magnetic fields
of any geometry, that endows the MRI with its special significance.
In fact, the MRI is merely one manifestation of much more general
behavior.  By imparting new degrees of freedom to a fluid, magnetic
fields allow free energy gradients to become sources of instability, with
important consequences for a variety of astrophysical systems: accretion
disks become turbulent when the angular velocity (not angular momentum)
decreases outward, and stratified dilute gases are destabilized when the
temperature (not entropy) decreases upwards \citep{b01}. 

The thermal destabilization caused by magnetic field is especially
noteworthy because it is engendered by what seems at first a purely
dissipative process: thermal diffusivity.   It is the extreme anisotropy
of the conductivity tensor parallel and perpendicular to the field lines
that lies at the heart of the instability.  By channeling heat only
along lines of magnetic force, small-amplitude ripples along initially
isothermal field lines grow into large fluid displacements parallel
to the temperature gradient.  This is not, as is sometimes thought,
analogous to classical double-diffusive instabilities, such as ocean
layer ``salt-fingering''\footnote{See Menou, Balbus, \& Spruit (2004)
for a true salt-fingering analogy involving the MRI.}.  Rather, it
depends wholly upon the properties of anisotropic conductivity.

In this paper we show that the anisotropy introduced into the viscous
stress tensor by a weak magnetic field sharply destabilizes dilute
astrophysical disks, even without Lorentz forces appearing in the fluid
equations.  By ``dilute,'' we mean the limit in which the ion Larmor
radius is small compared to a mean free path, which in turn is small
compared with the characteristic macroscopic length scales of the disk.
We shall refer to the resulting viscous stress tensor in this limit as
Braginskii viscosity \citep{b65}.   The Braginskii limit is appropriate
for interstellar and galactic disks (especially protogalactic disks,
cf. Malyshkin \& Kulsrud 2002).
A remarkable property of the instability is that in the absence of
Lorentz forces, when the viscous diffusivity much exceeds the resistive
diffusivity, rapid growth times are associated with arbitrarily high
wave numbers.  Except for isolated field geometries (e.g. precisely
azimuthal), there is no formal high wavenumber dissipation of the linear
magneto-viscous instability.  By way of contrast, for the ultra-weak
magnetic fields considered here, the wavelength of maximum growth in the
classical MRI would be strongly damped by an ordinary isotropic viscosity.
(Note that in real systems, at sufficiently high wave numbers, Lorentz
forces will ultimately stabilize.)

The magnetic stability of a dilute plasma may also be studied using
a plasma kinetic approach.  This offers rigor in a problem where it
is clearly of some benefit, but at a price of greater mathematical
complexity.  \citet{qdh02} analyzed the MRI in the collisionless regime,
and noted that the character of the instability changes when the pressure
tensor becomes significantly anisotropic, with growth rates in excess
of the classical MRI maximum.  In the current work, we have in essence
abstracted the anisotropic components of the pressure, and labeled
them as a viscosity tensor.  More recently, Sharma, Hammett \& Quataert
(2003) reanalyzed the kinetic problem using a Krook collision operator,
and showed that the collisional and collisionless behavior are on the same
branch of the dispersion relation.  In the short mean free path limit, the
pressure tensor associated with the equations used by Sharma et al. 2003
reduces to a scalar pressure plus the dominant parallel components of the
Braginskii viscosity (Snyder, Hammett, \& Dorland 1997).  The current
work, a purely gasdynamical treatment of the problem, affords relative
mathematical simplicity and a physically transparent interpretation.

In either its kinetic or gasdynamical guise, this vigorous instability
may be important in the early stages of magnetic field
amplification in disk galaxies, when densities are low, temperatures are
relatively high, and the field is likely to be very weak.  Quataert et
al.\ (2002) and Sharma et al.\ (2003) discuss applications to black hole
accretion flows of low radiative efficiency.

An outline of the paper is as follows.  Section 2 is a discussion of
the regime of the applicability of this work and a presentation of the
magnetic viscosity formalism.  Section 3 is the heart of the paper,
formulating and solving the problem, and checking the validity of the
results.  Section 4 is a discussion of the applicability of
our results to galactic magnetism, and section 5 summarizes our findings.

\section{Preliminaries}
\subsection{Gasdynamical Description of the Instability}

We begin with a physical description of the instability.  The process
is very simple.  In the leftmost diagram of figure 1 (a), we show
three representative azimuthal field lines, with increasing azimuth
toward the top.  The angular velocity gradient $\del\Omega$ is radial,
pointing from right to left, so that smaller angular velocities lie to
the right.  The angular momentum gradient, by contrast, runs from left
to right.  Initially, there is no viscous transport of angular momentum,
because the angular velocity gradient and field lines are orthogonal.

\begin{figure}
\epsscale {.90}
\plotone{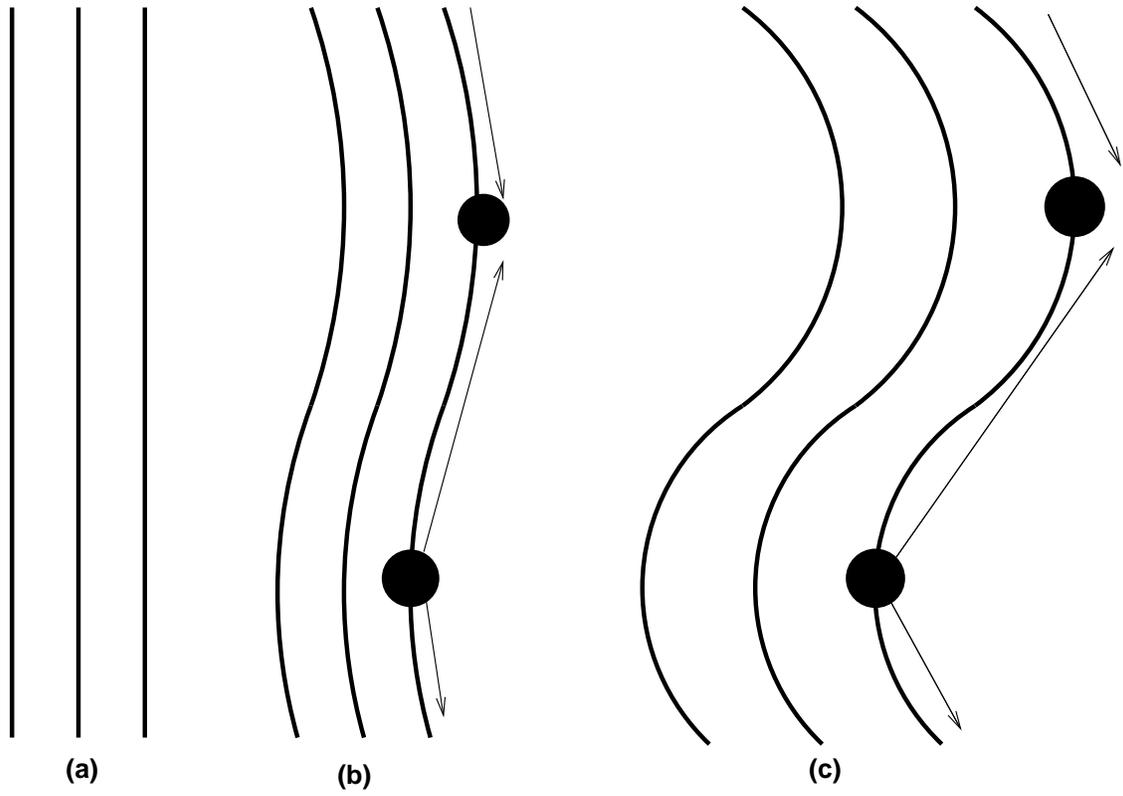}
\caption {Development of magneto-viscous instability
in three stages.  See text for details.}
\end{figure}

The middle portion of figure 1 (b) shows the field lines and two
fiducial fluid elements (dark circular disks) after a perturbation
has created a small oscillatory radial component.  Now there is a
component of $\del\Omega$ along the field lines, and viscous angular
momentum transport, represented by the thin arrows, is able to commence.
The rightward fluid element acquires angular momentum, while the leftward
element loses angular momentum.

In the rightmost portion of the figure (c), the loss of angular momentum
causes the element on the left to fall further leftward, towards orbits of
smaller angular momentum.  The element on the right, on the other hand,
acquires angular momentum, and moves toward the higher angular momentum
orbits that lie further to the right.  The critical point is that these
dynamical motions stretch the field lines yet more radially, allowing
yet more angular momentum to flow between the elements, and the process
runs away.

The instability blends different aspects of previously known
instabilities.  It works on the same dynamical principles as the MRI,
tethering fluid elements and transporting angular momentum between
them, but there is no magnetic tension here.  Similarly, it works
using the same gradient transport process seen in the magneto-thermal
instability, channeling diffusive transport along magnetic field lines
while simultaneously realigning them, but with angular momentum rather
than heat.  The result of all this, we shall see, is an extremely powerful
shear instability.

\subsection{Dilute Plasma Limit}

The presence of a magnetic field alters the form of the viscosity when
\beq\epsilon \equiv (\omega_{ci} \tau_{ci})^{-1}  \ll 1, \eeq
where $\omega_{ci}$ is the ion cyclotron frequency and $\tau_{ci}$ is the 
ion-ion mean collision time.
(This is tantamount to having the
mean free path much exceed the gyroradius.)  Under laboratory conditions,
this is normally considered to be
the large field strength limit, but it is very much the norm for interstellar
plasmas, even when the field is very weak.
Indeed, reference to \citet{s62} gives for a hydrogenic plasma
\beq
\omega_{ci} \tau_{ci} = \left(1.09\times 10^5\over n\right) {T_4^{3/2}B_{\mu G}
\over\ln \Lambda} ,
\eeq
where $n$ is the proton density in cm$^{-3}$, $T_4$ the kinetic
temperature in units of $10^4$ K, $B_{\mu G}$ is the magnetic field
in microgauss, and $\ln \Lambda$ is the Coulomb logarithm, $\sim 20$.
With $n \lesssim 1$ and $T_4 \gtrsim 1$, very weak fields are clearly
accommodated by the asymptotic regime $\epsilon \ll 1$.
The ion mean free path $\lambda_{mfp}$ and gyroradius $r_g$
are respectively
\beq
\lambda_{mfp} = 5.2\times 10^{11}\left(
20\over \ln \Lambda \right) \left(T_4^2\over n \right)
\ \mbox{cm}, \quad r_g = 1.3\times {10^8} \left(T_4^{1/2}\over B_{\mu G}
\right) \ \mbox{cm.}
\eeq

\subsection{Ideal MHD Limit}

Throughout this work, we ignore the effects of finite resistivity.
The ratio of the viscosity to resistivity is known as the magnetic Prandtl
number ${\cal P}$, and may be adapted from Balbus \& Hawley (1998):
\beq
{\cal P} = \left(T\over 10^4\right)^4\left(6.5\times 10^{10}\over n\right)
\left( 20\over \ln \Lambda\right)^2 .
\eeq
We may easily restrict the calculations to $T\gtrsim 10^4$ K and
$n \ll 10^{10}$ cm$^{-3}$, so that ${\cal P} \gg 1$, and resistivity will
be ignored.  Since these questions can be subtle however, we return
to this point {\em ex post facto} in \S 3.5.

\subsection{Magnetic Viscous Stress Tensor}

The theory of viscous transport in magnetized plasmas is presented by \citet{b65}.
The usual isotropic collisional viscous stress tensor can be written
\begin{equation}
\sigma_{ij} = - \eta W_{ij},
\end{equation}
where $\eta$ is the dynamical viscosity coefficient, and
\begin{equation}
W_{ij} = {\dd v_i\over \dd x_j} +{\dd v_j\over \dd x_i} -{2\over3}\delta_{ij}
\nabla \bcdot \bb{v}.
\end{equation}
This form applies to a set of Cartesian axes, $(i, j, k)$ being an
even permutation of $(X, Y, Z)$.  As usual, $\delta_{ij}$ denotes
the Kronecker delta function.  We note that the stress is traceless.
In the paper, we work exclusively in the Boussinesq limit, and shall
set $\del\bcdot\bb{v}=0$ in the above.  In \S 3.4, it is shown that the
Boussinesq limit is justified when the Reynolds number is large.

In the presence of a restricting magnetic field, the only component
of $\sigma_{ij}$ that remains unaffected is the momentum flux
along the magnetic line of force due to the gradient along
the field line.  Define a local Cartesian coordinate system
(the ``field frame'') $(\xb, \yb, \zb)$, chosen with
the magnetic field lying along the $\zb$ axis.
Then
\beq \label{sigzz1}
\sigma_{\zb \zb} = \sum_{i, j} b_i b_j \sigma_{ij},
\eeq
where the $b_i$ are components of the unit magnetic field vector
in an arbitrary Cartesian frame.
\citet{b65} shows that all other components of the magnetized
viscous stress tensor 
are smaller than $\sigma_{\zb \zb}$ by a factor of order $\epsilon$
or $\epsilon^2$, and are therefore ignored in this calculation.
The important exception are the two other
diagonal stress components, which to leading order in $\epsilon$ are identical
with one another.  Since
the stress must always be traceless, we have
\beq\label{sigxxyy}
\sigma_{\xb \xb} = \sigma_{\yb \yb} = -{1\over 2} \sigma_{\zb \zb}  .
\eeq
In vector notation, the $\zb\zb$ component of the stress is then given by
\beq\label{sigvec}
\sigma_{\zb \zb} = -2\eta \left[ (\bb{b}\bcdot\del)\bb{v}\right]\bcdot\bb{b}.
\eeq
To find the components of the magnetized viscous stress tensor in any other
locally Cartesian frame, the transformation law may be written
\beq
\sigma_{ij} = \sum_{i_b, j_b} 
(\bb{i}\bcdot \bb{i_b})\ (\bb{j}\bcdot\bb{j_b})\ \sigma_{{i_b} {j_b}}, 
\eeq
where once again the $b$ subscript denotes the magnetic field frame and
bold face quantities are unit vectors of the indicated component. Using
equation (\ref{sigxxyy}) for the nonvanishing diagonal stress tensor
components in the field frame, we find
\beq\label{sigij}
\sigma_{ij} = \sigma_{\zb\zb}\left[
( \bb{i}\bcdot \bb{Z_b})(\bb{j}\bcdot\bb{Z_b})
- {1\over 2}( \bb{i}\bcdot \bb{Y_b})(\bb{j}\bcdot\bb{Y_b})
-{1\over 2}( \bb{i}\bcdot \bb{X_b})(\bb{j}\bcdot\bb{X_b})
\right].
\eeq
Equations (\ref{sigvec}) and (\ref{sigij}) allow one to determine the
magnetic viscous stress in a frame suitable for working with fluid
variables, given the local field geometry.

\section{Formulation of the Problem}

We consider the stability of a disk in the presence of
a magnetic field, but with a field so weak that all dynamical
magnetic forces are negligibly small.  In contrast to the MRI,
we assume this ``magneto-anemic'' condition
is true not only for the equilibrium disk, but even
for small wavelength, WKB perturbations.  The sole effect of the
magnetic field is to restrict viscous transport in accordance
with prescription of \S 2.

The fundamental fluid equations used here are mass conservation
\beq
{\dd \rho\over \dd t} + \del\bcdot (\rho \bb{v} )= 0,
\eeq
the equation of motion,
\beq
\rho\left({\dd\ \over \dd t} + \bb{v}\bcdot\del \right)\bb{v} = 
-\del P -\rho \del \Phi -{\dd\sigma_{ij}\over\dd x_j},
\eeq
and the induction equation of ideal MHD,
\beq
{\dd\bb{B}\over\dd t} = \del\btimes(\bb{v}\btimes \bb{B}) .
\eeq
For reasons that will become clear, an
internal energy equation is not needed at this stage.

The equilibrium state is a differentially rotating disk, and we work in 
the usual cylindrical coordinate system, $R, \phi, Z$.  
The angular velocity is $\Omega(R)$, and we shall restrict
ourselves to a local analysis at the midplane.  Thus, we
may ignore buoyant forces, which depend upon gradients in 
pressure and entropy.  

In the equilibrium state, it is assumed that $\sigma_{ij} =0$.  Indeed,
the point of this calculation is to show that any development of viscous
transport along a field line is highly unstable.  The initial magnetic
field lines are wrapped around cylinders, and are unaffected by the shear.

We consider next small departures from the equilibrium flow.  Linearly
perturbed quantities are denoted by $\bb{\delta v}, \delta\sigma_{ij}$,
etc.  We work in the local WKB limit, with the space-time dependence of
all perturbed quantities given by $\exp(\gamma t + i\bb{k}\bcdot\bb{r})$.
Thus, $\gamma$ is a growth or decay rate if it is real, and an angular
frequency if it is imaginary.

The wavenumber $\bb{k}$ as well as the assumed constancy of $\gamma$
require some further explanation.  The wavenumber has radial,
azimuthal and vertical components $k_R, m/R, k_Z$ respectively.  Since we
are working in a local shearing system, the radial wavenumber $k_R$
will formally depend on time \citep{glb65, bh92}:
\beq\label{kR}
k_R(t) = k_R(0) - mt{d\Omega\over d\ln R} ,
\eeq
where $k_R(0)$ is the initial value of $k_R$.  For our present
purposes however, the time dependence of $k_R$ will prove irrelevant,
as the radial wavenumber disappears early in the analysis.  It is for
this reason that we may also assume a simple exponential time dependence;
in general the problem is more complex.

To evaluate $\delta\sigma_{ij}$, it follows from equation (\ref{sigij})
that
\beq
\delta \sigma_{ij} = \delta\sigma_{\zb\zb}\left[
( \bb{i}\bcdot \bb{Z_b})(\bb{j}\bcdot\bb{Z_b})
- {1\over 2}( \bb{i}\bcdot \bb{Y_b})(\bb{j}\bcdot\bb{Y_b})
-{1\over 2}( \bb{i}\bcdot \bb{X_b})(\bb{j}\bcdot\bb{X_b})
\right],
\eeq
since $\sigma_{\zb\zb}$ vanishes in the unperturbed state.  The terms
in square brackets may be evaluated for the equilibrium field geometry,
which we take to be
\beq
\bb{b} = \cos\theta\  \bb{{\hat \phi}} + \sin\theta\ \bb{{\hat Z}}.
\eeq
$\bb{{\hat \phi}}$, $\bb{{\hat Z}}$, and $\bb{{\hat R}}$
are unit vectors in the indicated  cylindrical
directions, and $\theta$ is the angle between the magnetic
field and the $\phi$ axis.  The local magnetic field axes are then
\beq\label{b}
\bb{Z_b} = \bb{b} = \cos\theta\ \bb{{\hat \phi}} + \sin\theta\ \bb{{\hat Z}},
\eeq
\beq
\bb{X_b} = \bb{{\hat R}}\btimes \bb{Z_b} = 
-\sin\theta\  \bb{{\hat \phi}} + \cos\theta\ \bb{{\hat Z}}, 
\eeq
and 
\beq
\bb{Y_b} = \bb{{\hat R}}.
\eeq
See fig.\ 2.

\begin{figure}
\epsscale {.90}
\plotone{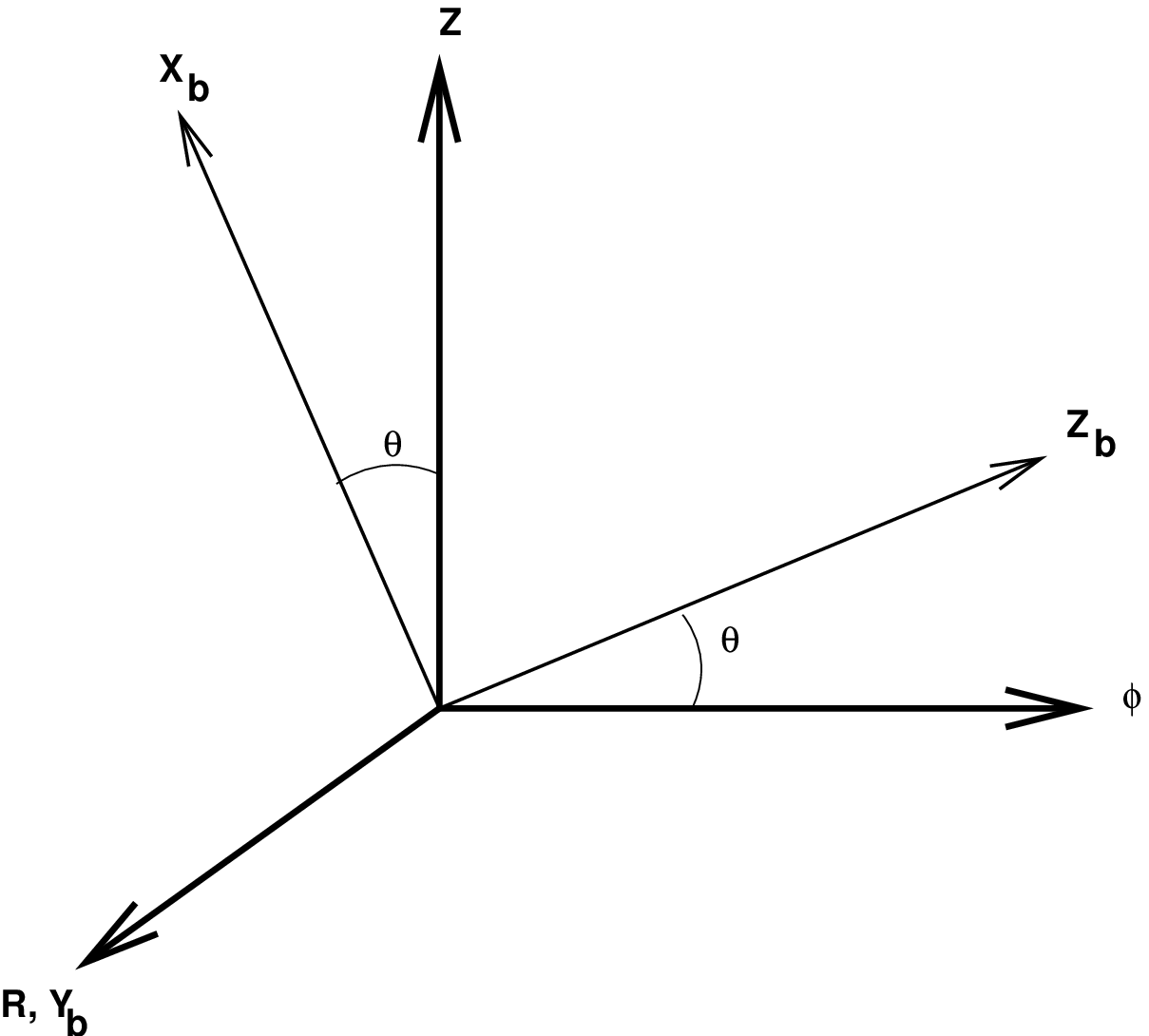}
\caption {Relative orientation of cylindrical $R,\phi, Z$ and magnetic
$X_b, Y_b, Z_b$ axes.  Magnetic field lies along $Z_b$, and
$R$ and $Y_b$ axes coincide.}

\end{figure}

The nonvanishing elements of $\delta\sigma_{ij}$ may now be calculated.
The diagonal elements are
\beq
\delta\sigma_{RR} = -{1\over 2}\delta\sigma_{\zb\zb}, \ 
\delta\sigma_{\phi\phi} = \left (\cos^2\theta - {\sin^2\theta \over 2}
\right) \delta\sigma_{\zb\zb}, \ 
\delta\sigma_{ZZ} = \left(\sin^2\theta -{\cos^2\theta \over 2}\right)
\delta\sigma_{\zb\zb},
\eeq
and the off-diagonal elements are
\beq
\delta\sigma_{\phi Z}=
\delta\sigma_{Z\phi}= {3\over 2}\cos\theta\ \sin\theta\ \delta\sigma_{\zb\zb} .
\eeq
We may evaluate $\delta\sigma_{\zb\zb}$ from its vector-invariant form
(\ref{sigvec}),
\beq
\delta\sigma_{\zb\zb} = -2 \eta 
\left(
\left[(\bb{\delta b}\bcdot\del)\bb{v}\right]\bcdot\bb{b}+
\left[(\bb{b}\bcdot\del)\bb{\delta v}\right]\bcdot\bb{b}+
\left[(\bb{b}\bcdot\del)\bb{v}\right]\bcdot\bb{\delta b}
\right) .
\eeq
Using $\bb{v}=R\Omega\bb{\hat \phi}$ and equation (\ref{b}), we find
\beq\label{sigzz2}
\delta\sigma_{\zb\zb} =-2\eta \left[ \delta b_R\, {d\Omega\over d\ln R}\, \cos\theta +
i(\bb{k}\bcdot\bb{b})(\bb{b}\bcdot\bb{\delta v})\right].
\eeq

\subsection{Linearized Equations}
The linearized equations are written in a coordinate system
that is shearing with the unperturbed flow.  The only
effects this has on the equations of motion are that
the time derivative must be Lagrangian,
\beq
{d\ \over dt} \equiv {\dd\ \over\dd t} + \Omega{\dd\ \over\dd\phi},
\eeq
and the radial partial derivative is replaced
by the $k_R(t)$ wavenumber in equation (\ref{kR})
\citep{bh92}.  The dynamical equations are
\beq
\bb{k}\bcdot\bb{\delta v} = 0,
\eeq
\beq
{d\, \delta v_R\over dt} -2\Omega \delta v_\phi +
{ik_R\over\rho}\left( \delta P -  {\delta\sigma_{\zb\zb}\over2}
\right)= 0,
\eeq
\beq\label{kap}
{d\, \delta v_\phi\over dt} + {\kappa^2\over 2\Omega} \delta v_R
+i{m\over\rho R} \left[
\delta P +\left({\cos^2\theta} - {\sin^2\theta\over2}\right)
 \delta\sigma_{\zb\zb}
\right]
+i{3k_Z  \over 2\rho} \sin\theta\, \cos\theta\, \delta 
\sigma_{\zb\zb} = 0,
\eeq
\beq
{d\, \delta v_Z\over dt} 
+{ik_Z\over\rho} \left[
\delta P +\left({\sin^2\theta } - {\cos^2\theta\over2}\right)
\delta\sigma_{\zb\zb} \right]
+i{3m\over 2\rho R} \sin\theta\, \cos\theta\, \delta 
\sigma_{\zb\zb} = 0.
\eeq
In equation (\ref{kap}), 
we have introduced the standard notation $\kappa$ for the epicyclic
frequency, which characterizes the response of
fluid element displacements in the $R\phi$ plane in a nonmagnetized
disk.  In terms of the angular velocity $\Omega(R)$,
\beq
\kappa^2 = 4\Omega^2 + {d\Omega^2\over d \ln R}.
\eeq

Notice that only the perturbed radial component of the magnetic field unit
vector $\delta b_R$ is required in $\delta\sigma_{\zb\zb}$.
This is easily obtained from the
radial induction equation:
\beq
{d \delta b_R\over dt} = i (\bb{k}\bcdot\bb{b}) \, \delta v_R.
\eeq
When the time dependence of the perturbations is a simple exponential,
this immediately gives
\beq\label{bR}
\delta b_R = {i (\bb{k}\bcdot\bb{b}) \over\gamma}\, \delta v_R,
\eeq
and
\beq
\delta\sigma_{\zb\zb} = -2\eta i (\bb{k}\bcdot \bb{b} ) \cos\theta \left[
{\delta v_R\over \gamma}{d\Omega\over d\ln R} +\delta v_\phi 
\right] -2\eta i (\bb{k}\bcdot \bb{b} ) \sin\theta\, \delta v_Z.
\eeq

\subsection {Axisymmetric Disturbances}

If there are both vertical and azimuthal field components present, the 
problem admits a relatively simple axisymmetric solution with $\bb{k}$
having only a vertical component.  This is also likely to be 
the most unstable mode, and therefore of greatest astrophysical
interest.  

With $\bb{k} = k_Z\bb{{\hat z}}$, mass conservation immediately
gives $\delta v_Z=0$.  The remaining dynamical equations are
\beq\label{kz1}
{d\, \delta v_R\over dt} -2\Omega \delta v_\phi  = 0,
\eeq
\beq\label{kz2}
{d\, \delta v_\phi\over dt} + {\kappa^2\over 2\Omega} \delta v_R
+{3\over 2\rho} ik_Z  \sin\theta\, \cos\theta\, \delta 
\sigma_{\zb\zb} = 0,
\eeq
\beq\label{check}
\delta P +\left({\sin^2\theta} - {\cos^2\theta\over2}\right)
\delta\sigma_{\zb\zb} = 0.
\eeq
Equation (\ref{check}) is one of vertical hydrostatic equilibrium.  In an
adiabatic gas, this constraint must be consistent with the Boussinesq approximation,
a point we revisit in \S 3.4.  

The problem decouples, and we are simply left with the two equations
(\ref{kz1}) and (\ref{kz2}) for $\delta v_R$ and $\delta v_\phi$.  
We may seek solutions with an exponential time dependence
of the form $\exp(\gamma t)$.  Then
\beq
{\delta\sigma_{\zb\zb}\over\rho} =
-2\nu ik_Z\sin\theta\, \cos\theta\, \left( {d\Omega\over d\ln R}
{\delta v_R\over\gamma} + \delta v_\phi\right),
\eeq
where we have introduced the kinematic viscosity,
\beq
\nu \equiv {\eta\over\rho}.
\eeq
Substituting this into equations (\ref{kz1}) and (\ref{kz2})
yields the dispersion relation
\beq\label{dispaxi}
\gamma^3 + (3\nu k_Z^2 \sin^2\theta \cos^2\theta)\gamma^2 + \kappa^2\gamma +
3\nu k_Z^2 \sin^2\theta \cos^2\theta\, {d\Omega^2\over d\ln R} = 0.
\eeq
This may be written
\beq
3\nu k_Z^2 \sin^2\theta\cos^2\theta = - {\gamma(\gamma^2 +\kappa^2)\over
\gamma^2 + {d\Omega^2/d\ln R}},
\eeq
from which it is clear there is an unstable branch ($\gamma > 0$)
of the dispersion relation if $\gamma+d\Omega^2/dR <0$.
The maximum growth rate is
\beq\label{gammax}
\gamma^2_{max}  = -{d\Omega^2\over d\ln R}.
\eeq
It is noteworthy that $\gamma$ may exceed the Oort A value $(1/2)
|d\Omega/d\ln R|$, a result emphasized by Quataert et al. (2002) for
a collisionless gas, who also recovered equation (\ref{gammax}) in the
appropriate limit.  We note here the interesting fact that the growth rate
given in (\ref{gammax}) will always exceed the Oort A value in any disk
that is locally stable by the Rayleigh criterion.  Equations (\ref{kz1})
and (\ref{kz2}) may be compared with equations (34) and (35) of Quataert
et al. (2002).  Note in particular the identification of the $\phi
Z$ component of the viscous stress tensor with the term proportional
to $\delta p_\parallel - \delta p_\perp$ on the right side of equation
(35) of Quataert et al. (2002).  In both the kinetic and gasdynamical
treatment, the azimuthal torque has no stabilizing radial counterpart,
which is responsible for the the resulting growth rate exceeding the
classical MRI Oort A value.

For a typical flat galactic rotation curve, the above gives $\gamma_{max}
= \sqrt{2}\Omega$.  This is an enormous rate: linear amplitudes grow a
factor of $7.2\times 10^3$ in one orbit, $5.2\times 10^7$ in two orbits.
We have of course neglected magnetic tension, which would ultimately be
a stabilizing influence, but not before it was a powerful destabilizing
influence in the form of the MRI.  A general treatment including the
dynamical effects of the field is deferred to a forthcoming paper.
But our finding certainly suggests that a kinematic treatment of field
amplification in galactic disks is at best questionable.  For plasma
kinetic treatments including Lorentz forces, see Quataert et al.
(2002) and Sharma et al. (2003).

\subsection {Azimuthal Field Instability}

The above analysis depended upon both axial and azimuthal field
components being present.  It is of interest to isolate the case of an
exactly azimuthal field ($\theta = 0$) and examine its stability properties.

We assume that the
wave vector is dominated by its axial component; however, in contrast
to the previous section, we do not set the azimuthal and radial components to
zero.  Rather, we work in the asymptotic limit
\beq
k_Z \gg k_R(0), m/R.
\eeq
The dynamical equations of motion in this case are
\beq\label{azi1}
{d\, \delta v_R\over dt} -2\Omega \delta v_\phi + {ik_R\over\rho}
\left( \delta P -{\delta\sigma_{\zb\zb}\over 2 }\right)= 0,
\eeq
\beq\label{azi2}   
{d\, \delta v_\phi\over dt} + {\kappa^2\over 2\Omega} \delta v_R
+i{m\over \rho R}\left( \delta P + \delta\sigma_{\zb\zb} \right) =0,
\eeq
\beq\label{azi3}
{d\, \delta v_Z\over dt} + {ik_Z\over\rho}
\left(\delta P -{\delta\sigma_{\zb\zb}\over 2 }\right)= 0.
\eeq
Mass conservation $\bb{k}\bcdot\bb{\delta v}=0$ implies that
$\delta v_Z$ is of order $1/k_Z$ relative to the planar velocity
components, and equation (\ref{azi3}) simplifies to
\beq
 \delta P-\delta\sigma_{\zb\zb}= 0.
\eeq
The remaining dynamical equations are
\beq
{d\, \delta v_R\over dt} -2\Omega \delta v_\phi =0,
\eeq
\beq
{d\, \delta v_\phi\over dt} + {\kappa^2\over 2\Omega} \delta v_R
+i{m\over\rho R}\left( {3\over 2}\delta\sigma_{\zb\zb} \right) =0.
\eeq
Once again, the radial wavenumber $k_R$ has dropped from the problem
(as has the dominant component $k_Z$),
and we may look for simple exponential time behavior of
the form $e^{\gamma t}$.  We find
\beq
\delta\sigma_{\zb\zb} = -2i \eta {m \over R}  \left[
{\delta v_R\over \gamma}{d\Omega\over d\ln R} +\delta v_\phi 
\right] .
\eeq
But 
this is exactly the problem we solved in the previous section, with the
substitution
\beq
k_Z \sin\theta\, \cos\theta \leftarrow m/R.
\eeq
We obtain exactly the same dispersion relation, 
\beq\label{dispazi}
\gamma^3 + 3\gamma^2\nu m^2/R^2 + \kappa^2\gamma +
3\nu (m^2/R^2){d\Omega^2\over d\ln R} = 0,
\eeq
and thus exactly the same
maximum growth rate, $|d\Omega^2/d\ln R|$.  

\subsection {Validity of the Boussinesq Approximation}

The Boussinesq approximation was very important for
simplifying the preceding analyses.  We review its validity
here.

The key point is that the perturbed pressure is of order
\beq
\delta P \sim \delta\sigma_{\zb\zb} \sim 
(\bb{k}\bcdot \bb{b})\eta \delta v_\phi,
\eeq
since $\delta v_R$ and $\delta v_\phi$ are comparable and the
growth rates are of order $\Omega$.  Assuming that any
radial entropy gradient is very small, 
\beq
{\delta\rho\over\rho} \sim {\delta P\over P}\sim{\delta P\over \rho c_S^2},
\eeq
where $c_S$ is the isothermal sound speed.  
Hence
\beq
{\delta\rho\over\rho} \sim {\nu (\bb{k}\bcdot \bb{b})\over c_S}
{\delta v_\phi\over c_S}.
\eeq
The Boussinesq approximation requires
\beq
{d\ln \rho\over dt} \sim \Omega {\delta\rho\over\rho} \ll k\delta v,
\eeq
where $k$ and $\delta v$ are respectively 
characteristic values for the wave number and perturbed velocity.
With $k\sim \bb{k}\bcdot \bb{b}$, this requirement simplifies to
\beq
{\nu\over (c_S^2/\Omega)} \ll 1.
\eeq
The quantity on the left is $1/Re$, the inverse of the Reynolds number.
The Boussinesq limit is therefore generally appropriate
for our problem in the limit of large Reynolds number. 

\subsection{Effect of Resistivity}

A finite resistivity will always be present, directly affecting the
dynamics because of its role in modifying the constraint of field
freezing.  The question arises as to whether resistivity may generally
be ignored at all wavenumbers if the viscous diffusivity $\nu$
much exceeds the resistive diffusivity $\eta_B$.
(Unlike the transformation of the viscous stress tensor, the parallel
and transverse resistivities do not differ profoundly [Spitzer 1962],
and we shall ignore the distinction here.)

The presence of Ohmic resistance alters equation (\ref{bR})
to
\beq
b_R = {i (\bb{k}\bcdot \bb{b})\over \gamma + \eta_B k^2}\, \delta v_R,
\eeq
where $k^2$ is the magnitude of the wavenumber.  In both of our
worked examples, $k$ may be taken as $k_Z$.  

It is a straightforward exercise to rework the dispersion formula
using the above for $b_R$.  One obtains for the case of
axisymmetric disturbances, 
\begin{eqnarray}
\gamma^3 + \gamma^2k_Z^2(\eta_B + 3\nu \sin^2\theta\cos^2\theta)
+\gamma (\kappa^2 +3\eta_B\nu k_Z^4\sin^2\theta \cos^2\theta)+&&\nonumber\\
+ k_Z^2\left(\eta_B  \kappa^2 + 3 \nu\sin^2\theta\cos^2\theta
{d\Omega^2\over d\ln R}\right) = 0.&& { }
\end{eqnarray}
As an equation for $k_Z^2$ this reads
\begin{eqnarray}
3k_Z^4(\gamma \eta_B \nu\sin^2\theta \cos^2\theta) 
+ k_Z^2  \left[ \eta_B(\gamma^2 +\kappa^2)+ 3 \nu
\sin^2\theta \cos^2\theta
(\gamma^2 +d\Omega^2/d\ln R)
\right]+ &&\nonumber\\
+\gamma(\gamma^2 + \kappa^2) = 0.&&
\end{eqnarray}
For $\gamma>0$,
positive solutions for $k_Z^2$ exist only if the coefficient of 
$k_Z^2$ is negative, which, in the limit
$\nu \gg\eta_B$, requires $d\Omega^2/dR<0$ as before.  
The quadratic discriminant in the solution for $k_Z^2$  must also 
be positive.  If we set 
\beq
\gamma^2 + {d\Omega^2\over d\ln R}  = \epsilon\, {d\Omega^2\over
d\ln R},
\eeq
and assume that $\epsilon \ll 1$, the condition for the positivity
of the discriminant is easily worked out.  We find
\beq
\epsilon^2 > {64\eta_B\over 3\nu\sin^2(2\theta)} \left|d\ln\Omega^2\over d\ln R\right|^{-1}.
\eeq
Provided that $\theta$ is not too small, 
this is consistent with our small $\epsilon$ assumption, and shows that
a small resistivity will not strongly alter the conclusions of \S 3.2.

For the case of an azimuthal magnetic field, the above breaks down
and a separate analysis is needed.  There is, already at the start,
a small parameter in the problem: $m/k_Z R$. The requirement that that
resistivity be negligible is not satisfied merely by $\eta_B\ll\nu$,
but by the tighter restriction
\beq\label{ineq}
\eta_B \ll 3\nu (m/k_ZR)^2.
\eeq
This is seen explicitly in the dispersion relation:
\begin{eqnarray}\label{etax}
\gamma^3 + \gamma^2\left(\eta_B k_Z^2 + 3 \nu {m^2\over R^2}\right)
+\gamma\left( \kappa^2 + 3\eta_B\nu k_Z^2 {m^2 \over R^2} \right)+&&\nonumber\\
+\eta_B k_Z^2 \kappa^2 + 3\nu {m^2\over R^2} {d\Omega^2\over d\ln R} = 0.&&
\end{eqnarray}
There can be no instability unless the constant term is negative, which requires
the inequality (\ref{ineq}) to hold.  The asymptotic regime of validity
for this condition is thus
\beq\label{calP}
1\ll {k_Z R \over m} \ll  (3 {\cal P})^{1/2}.
\eeq
There are further restrictions, however, if the maximum
growth rate is not to depart significantly from its value
in equation (\ref{gammax}).  If we regard the dispersion
relation
as an equation for $m^2/R^2$, it may be shown that
it has a well-defined
solution for $\gamma\rightarrow\gamma_{max}$, only if
\beq
\eta_B k_Z^2 \ll 
\left|d\Omega^2/d\ln R\right| .
\eeq
Similarly, if we regard the dispersion relation as an equation $k_Z^2$,
the restriction is found to be
\beq
3\nu (m/R)^2 \gg 
\left|d\Omega^2/d\ln R\right| .
\eeq
The last three equations may be combined to give the final form
for the asymptotic domain of the axisymmetric dispersion relation
(\ref{etax}) in which our solution for
$\gamma_{max}$ remains unaffected:
\beq
1\ll {k_Z R \over m} \ll \left(R\over m\right)
\left|{1\over \eta_B} {d\Omega^2\over
d\ln R} \right|^{1/2} \ll  (3 {\cal P})^{1/2}.
\eeq

\section{Discussion: Origin of Galactic Magnetic Fields.}

The instability presented in this paper is generic and powerful.  Its most
rapidly growing behavior is exhibited at large vertical wavenumbers, and
even in the presence of finite resistivity, it is not easily quenched.
Long radial and azimuthal wavelengths are highly unstable, imparting
large scale coherence in the disk plane already in the linear stages
of instability.  In this section we discuss the possibility that this
mechanism could be an important part of the process that gives rise to
galactic magnetic fields.

Generally, galactic magnetic amplification processes are divided into
two principal categories \citep{beck96}: (1) differential wind-up of a
primordial field (Kulsrud 1986); and (2) dynamo amplification of a similar
seed field.  The latter category itself consists of at least two very
distinct mechanisms; (2a) a classical $\alpha\Omega$ dynamo (Parker 1979),
and (2b) small scale turbulent amplification (Schekochihin et al. 2004).
The weak field limit of these mechanisms are thought to be kinematic
in nature.  It is difficult to see how the process outlined in this paper
could be unimportant, at least quantitatively, to any of these processes.

Simple wind-up of a seed field by differential rotation fails to
incorporate properly the true dynamics of either the MRI or the
magneto-viscous instability described here.  Differential rotation and
essentially any field geometry is highly unstable, resulting in large
radial motions and ultimately MHD turbulence.  The shearing of radial
fields of course readily occurs, but it is not the primary amplification
process of a differentially rotating system.

Traditional $\alpha\Omega$ galactic dynamo models rely on the presence of
favorable properties of interstellar turbulence to generate the required
mean helicity.  Curiously, these theories of magnetic field amplification
generally do not incorporate weak-field MHD instabilities, despite the
latter's obvious potential benefits (e.g. Brandenburg et al. 1995; Hawley,
Gammie, \& Balbus 1996).  Conversely, without something like the MRI or
the magneto-viscous instability, it is not so obvious that turbulence
will generally amplify the field, at least at low to moderate values of
${\cal P}$.  An example of dissipative behavior is seen in a simulation
of Hawley et al.\ (1996).  The combination of hydrodynamical shear-layer
turbulence plus a magnetic field lead not to dynamo activity, but to field
energy loss.  This result emerged despite the fact that calculation
was done not in the kinematic limit, but fully in the MHD regime.
However, the combination of local Coriolis, tidal, and Lorentz forces
dramatically altered the dynamo properties of the ensuing turbulence:
rapid and significant field amplification was observed, driven by the
MRI turbulence, in a hydrodynamically stable background.

More recently, the notion that non-helical homogeneous small scale
turbulence may play a key role in galactic dynamos, particularly in the
early stages, has been investigated in a series of numerical simulations
(Schekochihin et al. 2004; see also Zeldovich et al. 1984, Kulsrud \&
Anderson 1992), which include a scalar viscosity
and study the non-kinematical regime.  The idea
is that certain magnetic field configurations (termed ``winning''
by Schekochihin et al.) align themselves smoothly with the stretching
direction of the strain tensor of the turbulent flow, but fluctuate along
the corresponding null axis, so that the work done on the field by fluid
element stretching is not undone by relaxation.  A sort of turbulent
ratchet thereby ensues, growing the field with enormous efficiency at
small scales.  Numerical simulations of homogeneous white noise forcing
conducted by Schekochihin et al. 2004 resulted in exponential field
amplification, but, interestingly, only in the regime of large ${\cal P}$.
What the relationship is between these small scale structures and a 
galactic scale field has yet to be established.  

Further discussion of the pros and cons of this mechanism would take
us too far afield, but we may note that the generation of coherent
magnetic field structure on scales larger than the disk scale height
requires time scales at least as long as $1/\Omega$.  And while it
may well be that intrinsic interstellar turbulence plays a role in
the amplification of galactic magnetic fields, we emphasize here a
phenomenon that is certainly unavoidable: differential rotation.  In fact,
there is no reason to restrict oneself to galactic radius scales:
if sub-galactic, differentially rotating turbulent vorticies are also present, 
magneto-viscous modes should be seen on these scales as well.
The key point is that a kinematic approach will miss this process,
which is intrinsically MHD.

One compelling origin of the seed field for the galactic amplification
process is the stars themselves (Biermann 1950, Rees 1993), in which
both battery processes and a truly powerful, rapid dynamo are likely to
be present.  A 0.1 G azimuthal surface field diluted in a stellar wind 
to interstellar scales would give rise to a seed field of $\sim 2\times
10^{-9}$G.
With $T\simeq 10^4$ K, $n\simeq 1$ cm${}^{-3}$, one finds
$$
\omega_{ci}\tau_{ci} \simeq 10, \quad {\cal P} \sim 10^{11}
$$
and the regime of the magneto-viscous instability is valid.
The ratio of gas pressure to magnetic pressure is
$3.5\times 10^7$, an extremely weak field by this measure.
The linear amplification factor per orbit of the magnetic energy,
as noted in \S 3.2, is huge: $5.2\times 10^7$.
The makings of an MHD dynamo would seem to be present.

If a combination of magneto-viscous and magneto-rotational processes is
to be a viable candidate for galactic field production, it needs to be
shown that (1) the bulk of the field energy emerges in the largest scales;
and (2) the saturated field energy density can grow to levels comparable
to the thermal pressure.  Definitive answers will require a numerical
treatment, but in the meantime the following points may be considered.

A magnetic energy spectrum dominated by the largest scales appears
to be a universal outcome of MRI simulations (e.g. Hawley, Gammie, \&
Balbus 1995), despite the fact that the most unstable modes occur at
scales much smaller than the scale height.  In such calculations, the
available dynamical range is limited.   To the extent it can be measured,
however, the inertial range is Kolmogorov-like in the magnetic energy.
In hydrodynamics, Kolmogorov scaling for the kinetic energy power
spectrum is universal, if the dissipation rate per unit volume is the
sole constant characterizing the cascade, as often seems to be the case.
Universal processes may also be at work in MHD turbulence (Kraichnan
1963; Goldreich \& Sridhar 1997).  An important caveat, however, is that
most MHD turbulence studies have been based on wave-wave interactions (which
may be of secondary importance in linearly unstable rotating systems),
and they have yet to address the role of a magnetized viscosity.

Indeed, a magnetized viscosity may be crucial to understanding why
interstellar fields are thermal strength (or even slightly above),
whereas all numerical MRI simulations to date have yielded subthermal
saturated field strengths.  It has been argued (Balbus \& Hawley 1998)
that the outcome may be well be sensitive to ${\cal P}$.  The point is
that large scale reconnection proceeds relatively easily in simulations
when the resistive scale is comparable to or larger than the viscous
scale.  In ``ideal MHD,'' both scales are grid based.  But matters are
likely to be very different if the viscous scale is, say, five orders of
magnitude larger.  In that case, enormous viscous stresses would occur in
the course of setting up a reconnection front, and prevent its formation.
With reconnection stifled, MHD turbulence could amplify the field to
its natural dynamical limit: the thermal energy density.  Beyond this
point, buoyant effects would make it difficult for a suprathermal field
to remain in the disk and be further amplified (Parker 1979).

Large Prandtl number simulations have in fact recently begun (Schekochihin
et al. 2004), and treatments of the anisotropic Braginskii viscosity
and conductivity have yet to be attempted.  With the resistivity scale
hidden below the viscous dissipation scale, there is no effective small
scale sink for the magnetic field, and is therefore not surprising to note
that Schekochihin et al.\ find that the magnetic energy spectra increases
with wave number on subviscous scales, before it is ultimately cut-off.
By way of sharp contrast, the ideal MRI simulations noted above find a
monotonically decreasing energy spectrum for the magnetic field (Hawley,
Gammie, \& Balbus 1995).

The magneto-viscous instability is a powerful and general mechanism to
amplify weak magnetic fields in galaxies, or in hot dilute plasmas more
generally, and is worthy of more detailed study.  Of particular
interest would be a suite of numerical simulations designed to isolate
the differences between high Prandtl number turbulent dynamos relying
on MHD and differential rotation, those based on random forcing only,
and those containing both.

\section{Summary}
When the ion Larmor radius is less than its collisional mean free path,
the form of the viscosity is altered.  Crudely speaking, angular momentum
is transported only along magnetic lines of force.  This causes conjoined
fluid elements in a differentially rotating system to separate radially,
dragging field lines with them.  This aligns the field lines along the
angular velocity gradient, causing an increase in the diffusive angular
momentum transport, and the process is highly unstable.  The instability
has previously been studied in its plasma kinetic guise (Quataert et
al. 2002; Sharma et al. 2003), and we find quantitative agreement in
areas of overlap with this approach.  The magneto-viscous instability
does not rely directly upon magnetic stresses; the field serves merely
to channel angular momentum transport along its lines of force.  This is
effective even at exceedingly weak magnetic field strengths.  The maximum
growth rate of the instability is given by equation (\ref{gammax}),
and exceeds the Oort-A value of any disk stable by the hydrodynamical
Rayleigh criterion.  Numerical simulation of the nonlinear resolution
of the instability promises to be a challenging problem.  Preliminary
studies on the related magneto-thermal instability, have, in fact, begun
(J. Stone 2004, private communication).

The long mean free path regime is appropriate to nonradiative accretion
flows around black holes (Quataert et al. 2002; Sharma et al. 2003), or
to the interstellar medium of disk galaxies, our interest in this work.
The magneto-viscous instability is thus a candidate for amplifying very
small galactic seed fields into thermal strength fields.  The turbulent
MHD power spectrum, if it may be extrapolated from the moderately
resolved, relatively low Prandtl number regime of previous examples,
would evidence most of the power on large scales.  On the other hand,
the large Prandtl number regime, which characterizes the interstellar
medium, ensures that the resistive dissipation scale is much smaller than
the viscous scale.  This may make it possible to grow thermal strength
magnetic fields, despite the subthermal fields obtained in numerical
MRI simulations, which correspond to the opposite Prandtl regime.  But,
at the same time, it is possible that the high Prandtl number regime may
skew the power spectrum toward smaller scales (Schekochihin et al. 2004).

A very general gasdynamical investigation of the instability, including
its behavior for arbitrary wavenumber directions, and the dynamical
stress of field line tension, will be presented in a forthcoming paper
(Islam \& Balbus 2004, in preparation).

\section*{Acknowledgements}

I would like to thank the referees G.W.\ Hammett and E.\ Quataert for an
extremely constructive and thoroughgoing review.  I am also grateful
to T. Islam for useful discussions, to S. Cowley, J. Stone, and
A. Schekochihin for comments on an earlier draft of this work, and to
the Ecole Normale Sup\'erieure for providing financial support and a
stimulating atmosphere, which fostered the completion of this work.
Support from NASA grants NAG5--13288 and NAG5--9266 is 
gratefully acknowledged.




\end{document}